\documentclass[10pt,final]{IEEEtran}
\usepackage{subcaption}
\usepackage{cite}
\usepackage{amsmath,amssymb,amsfonts}
\usepackage{graphicx}
\usepackage{textcomp,nicefrac}
\usepackage{import}
\usepackage{graphicx}
\usepackage{enumitem}
\usepackage{tabularx}
\usepackage[table]{xcolor}
\usepackage{multirow}
\usepackage{makecell}
\usepackage{url}
\usepackage{xcolor}
\usepackage{tikz}

\newcommand{\copyrighttext}{%
  \footnotesize \textcopyright 2022 IEEE. Personal use of this material is permitted.
  Permission from IEEE must be obtained for all other uses, in any current or future 
  media, including reprinting/republishing this material for advertising or promotional 
  purposes, creating new collective works, for resale or redistribution to servers or 
  lists, or reuse of any copyrighted component of this work in other works. 
  URL: \url{https://ieeexplore.ieee.org/document/9733910} DOI: 10.1109/TNS.2022.3159175}

\newcommand{\copyrightnotice}{%
\begin{tikzpicture}[remember picture,overlay]
\node[anchor=south,yshift=8pt] at (current page.south) {\fbox{\parbox{\dimexpr\textwidth-\fboxsep-\fboxrule\relax}{\copyrighttext}}};
\end{tikzpicture}%
}

\usepackage{pgfplots}
\usepackage{pgfplotstable}
\pgfplotsset{width=\linewidth,compat=1.9}
\usepackage{tikz}
\usetikzlibrary{positioning}

\pgfplotstableread{
x         y             ydev                z                    zdev

1	1.5574598121643066  0.6163553160805103  1.69436104774475	0.735551063758264
5	5.291354146003723	1.5910975786382864  5.30580804824829	1.55173184342134
10	7.549356622695923	1.6542072739106786  8.26171613454819	1.4565497930344
20	11.084053182601929	1.2619461197749733  11.6348831033707	1.17110389009444
50	15.553806097507477	1.0228372006662356  16.5363303399086	0.866261229081277
100	18.990108456611633	0.7544444761898365  20.8252662110329	0.758195765021348
200	23.12503336429596	0.7217902688196178  26.4978225088119	0.735688622443957
500	30.996901905536653	0.7804866082100621  39.4189419579506	0.753875823418026
1000	42.111338307857515	0.9707527124279068  58.5564357995987	1.19845991897587

}{\mytable}

\def\BibTeX{{\rm B\kern-.05em{\sc i\kern-.025em b}\kern-.08em
T\kern-.1667em\lower.7ex\hbox{E}\kern-.125emX}}
\begin{document}

\title{Contextual Isotope Ranking Criteria for Peak Identification in Gamma Spectroscopy Using a Large Database}

\author{Alexis Aguilar-Arevalo,
Xavier Bertou,
Carles Canet,
Miguel A. Cruz-Pérez,
Alexander Deisting,
Adriana Dias,
Juan Carlos D'Olivo,
J. Francisco Favela-Pérez,
Estela A. Garcés,
Adiv González Muñoz,
Jaime Octavio Guerra-Pulido,
Javier Mancera-Alejandrez,
Daniel José Marín-Lámbarri,
Mauricio Martínez-Montero,
Jocelyn Monroe,
Sean Paling,
Simon Peeters,
Paul R. Scovell,
Cenk T\"urko\u{g}lu, 
Eric Vázquez-Jáuregui,
Joseph Walding

\thanks{This work is supported by the STFC Global Challenges Research Fund
(Foundation Awards, Grant ST/R002908/1 and Translation Awards, Grant
EP/T015586/1), DGAPA
UNAM grants PAPIIT-IT100420 and PAPIIT-IN108020, CONACyT grants CB-240666 and A1-S-8960. M. A. Cruz-Pérez acknowledges the support of ``Programa de Posgrado en Ciencias de la
Tierra at Universidad Nacional Autónoma de México.''}

\thanks{A. Aguilar-Arevalo, J. C. D'Olivo, J. F. Favela-Pérez, J. O. Guerra-Pulido, and M. Martínez~Montero are with Instituto de Ciencias Nucleares, Universidad Nacional Autónoma de México, Coyoacán, CDMX, México, C.P. 04510 (e-mails: francisco.favela@correo.nucleares.unam.mx, jaime.guerra@correo.nucleares.unam.mx, \\mauricio.martinez@nucleares.unam.mx).} 

 \thanks{X. Bertou and J. F. Favela-Pérez are with Centro Atómico Bariloche,  CNEA/CONICET/IB, Bariloche, Argentina.}
 
 \thanks{A. Deisting, A. Dias, J. Monroe, and J. Walding are with Royal Holloway, University of London, Egham Hill, United Kingdom.}
 
 \thanks{E. A. Garcés, A. González Muñoz, D. J. Marín-Lámbarri, and E. Vázquez-Jáuregui are with Instituto de Física, Universidad Nacional Autónoma de México, A. P. 20-364, México D. F. 01000, Mexico.}
 
\thanks{C. Canet and M. A. Cruz-Pérez
 are with Centro de Ciencias de la Atmósfera. Universidad Nacional Autónoma de México, Ciudad Universitaria, Coyoacán, 04510, Ciudad de México, Mexico.}

\thanks{J. Mancera-Alejandrez is with Facultad de Ingeniería, Universidad Nacional Autónoma de México, México.}

\thanks{S. Paling and P. R. Scovell are with Boulby Underground Laboratory, Boulby Mine, Saltburn-by-the-Sea, United Kingdom.}

\thanks{S. Peeters and
C. T\"urko\u{g}lu are with Department of Physics and Astronomy, University of Sussex, Brighton, United Kingdom.}

\thanks{C. Canet is also with Instituto de Geofísica, Universidad Nacional Autónoma de México, Ciudad Universitaria, Coyoacán 04510, Ciudad de México, México.
}

}

\maketitle
\copyrightnotice
\begin{abstract}
Isotope identification is a recurrent problem in $\gamma$ spectroscopy
with high purity germanium detectors. 
In this work, new strategies are introduced to facilitate this type of analysis.
Five criteria are used to identify the parent isotopes making a query on a large
database of $\gamma$-lines from a multitude of isotopes producing an output list
whose entries are sorted so that the $\gamma$-lines with the highest chance of
being present in a sample are placed at the top. A metric to evaluate the
performance of the different criteria is introduced and used to compare them. Two
of the criteria are found to be superior than the others: one based on fuzzy
logic, and another that makes use of the $\gamma$ relative emission probabilities. A
program called \texttt{histoGe} implements these criteria using a SQLite
database containing the $\gamma$-lines of isotopes which was parsed from WWW Table
of Radioactive Isotopes. \texttt{histoGe} is Free Software and is provided along
with the database so they can be used to analyze spectra obtained with
generic $\gamma$-ray detectors.

\end{abstract}


\section{\label{sec:level1}Introduction}
 
\IEEEPARstart{G}{amma} ray spectroscopy commonly uses High Purity Germanium Detectors (HPGe) to acquire the energy spectra of samples with a varied isotopic composition. Analyzing and identifying the isotopes present in the sample is a challenging problem given the multitude of gamma lines from which to choose. Thus, computational tools are needed to analyze a spectrum and extract useful information \cite{gilmore2011practical,fagan2012statistical}. 
Many programs have been written 
to make these calculations in the past \cite{kalfas2016spectrw,simonits2003hyperlab,wasim2010gammalab,arnold20052002,nielsen1998intercomparison,blaauw19971995,zahn2015evaluation,Guilherme2015}. However, just a few provide the capability to identify
isotopes from a given spectrum. Some previous attempts have
been made to perform that identification using a relational database,
for example: Hyperlab \cite{simonits2003hyperlab} has a functionality to make
a ``graphical iteration process'' followed by a procedure to solve
iteratively an ``identification matrix'', however, no documentation
about the method nor reports on its identification accuracy are
provided for this software. GammaLab \cite{wasim2010gammalab}
uses a database called ``NUCDATA'' which includes information ``for
quick calculations for 408 radioisotopes'' \cite{wasim2002nucdata},
however, it is limited in extension in comparison to other well-known
databases \cite{nucData,IAEALiveChart,nndc} and it does not present a
study about its accuracy and its real capability of identifying
isotopes in a sample. ASPRO-NUC \cite{kolotov1995software} has a wide
set of spectral analysis tools: peak search,
deconvolution, background line and simulation, spectrum smoothing,
among others. It also has algorithms to identify peaks using a
database of 45,000 $\gamma$-lines corresponding to 2200
radionuclides. However, the authors recognized that identification in this
way ``is hardly possible'' and they opted to develop an ``actual
isotope library''. Sandia National Laboratories provide two programs for 
assisting with analyzing spectral information from nuclear radiation.  \texttt{InterSpec} \cite{interspec} provides 
multipeak fitting isotope identification capabilities as well as activity 
calculation considering shielding from a variety of materials. The other one
is called Peak Map \cite{peak-map} which is written in C\#,
it considers a set of parameters such
as distance from the mean and half-life penalization (among others) to 
assign a score to the $\gamma$-lines and sorts with respect to that before
displaying the candidate $\gamma$-lines. 

There are many areas in which $\gamma$-spectroscopy is applied, e.g., high energy physics 
research, environmental sciences, and food contamination. An important application lies
in the control of radioactive materials crossing borders
around the world, where there is a need to have
instruments and methods to identify radioactive nuclei that
are potentially harmful even in small amounts. While automated analysis
for this problem is desirable, a study from 2007 
concluded that a ``secondary analysis of spectra
by a trained spectroscopist is frequently necessary'' to identify
isotopes through their $\gamma$-lines \cite{sullivan2007evaluation}. Since then, new techniques
have been explored to make automated isotope
identification more reliable independently of the field of application,
to mention a few: swarm optimization
\cite{shahabinejad2018analysis}, Fischer linear discriminant analysis
\cite{boardman2012gamma}, Bayesian statistics approach
\cite{sullivan2015validation,kim2020quantification}, neural networks
(NN)
\cite{kamuda2019automated,lagari2017rbf,kamuda2017automated,kim2019multi,varley2015development,kamuda2020comparison,blazquez2015classification},
hybrid fuzzy-genetic algorithms \cite{alamaniotis2015hybrid}. 
Some of these methods have been used to perform automated
peak identification \cite{stinnett2016automated}. There are also
developments, using the GEANT4 toolkit, for providing training
data to machine learning models \cite{Turner_2020}.
Herein, new methods have been proposed, implemented, and tested using the histoGe code \cite{histoGeRepo}, a Free Software (GNU Public License \cite{gplv3}) with many features that are described in its User's Manual \cite{histoGeUserManual}.

This work deals with the $\gamma$-line identification
problem in a manner similar to what a
search engine does when presenting the results of a query. It assigns
a numerical value called ``rank value'' (RV) 
to each candidate $\gamma$-line
that may explain the presence of the peak in a spectrum.

Five criteria 
to identify peaks are presented, some of them are based on simple counting
while others use more complex calculations, e.g., Mamdani's fuzzy inference system (FIS). 
Known $\gamma$ spectra were identified using the proposed criteria and their performance was evaluated using an \textit{ad hoc} metric.

This work has been developed as an effort to
support the research and educational activities
that will be carried out at the 
Laboratorio Subterráneo de Mineral del Chico
(LABChico), which will be located inside a decommissioned silver mine at the
Comarca Minera, Hidalgo, México, inside the UNESCO Global
Geopark \cite{comihi}. LABChico will host HPGe to conduct studies of low
radioactivity in water, soils, and products intended for human
consumption, aiming to develop techniques to signal the presence of
lead in drinking water. At the same time, it will serve as a training
hub for students and researchers interested in radiation detectors,
techniques for particle and astroparticle physics experiments,
geology and mine engineering, among other areas. The \texttt{histoGe}
computational software was developed as an effort to facilitate
$\gamma$-ray spectroscopy and to identify isotopes from recorded
spectra inside the laboratory. Technical details about its implementation and capabilities, how to use it, and its database of \texttt{histoGe} are described in the user's manual \cite{histoGeUserManual}.

This paper is organized as follows: in section \ref{sec:genProc}, the
methodology used to identify isotopes and a brief explanation of the operation of the
program and some key concepts are presented; in section \ref{sec:crit}, five
criteria to find the most suitable $\gamma$-lines that can be responsible for
the peaks observed in the spectra are described; in section \ref{sec:experimental}, the experimental setup is described; in section
\ref{sec:results_complete} four cases are studied: one is an example of how the
general method works, the second one analyzed the spectra obtained with
point-like radioactive sources through the Nuclear Sciences Institute HPGe 
(ICN-HPGe) detector, the third example is the analysis of spectra of some
samples of rocks and water and the fourth example presents the
identification of isotopes using a spectrum taken from the literature.
The last section shows the conclusions of this work.

\section{\label{sec:genProc} Methodological approach}
The \texttt{histoGe} software is written in
Python~3~\cite{10.5555/1593511} and can run
practically in any of the mainstream operating 
systems available nowadays. The basic process of line sorting used in \texttt{histoGe} is depicted in Figure 1. Energy query 
ranges are determined by either processing an experimental spectrum to find peaks, or by hand. \texttt{histoGe} uses a 
Savitzki-Golay filter \cite{savGolayFilter} to smooth out spectra, detect the peaks 
and generate these query ranges. The width of the 
query ranges contains information about the resolution of 
the detector.

\begin{figure}
    \centering
    \includegraphics[scale=0.5]{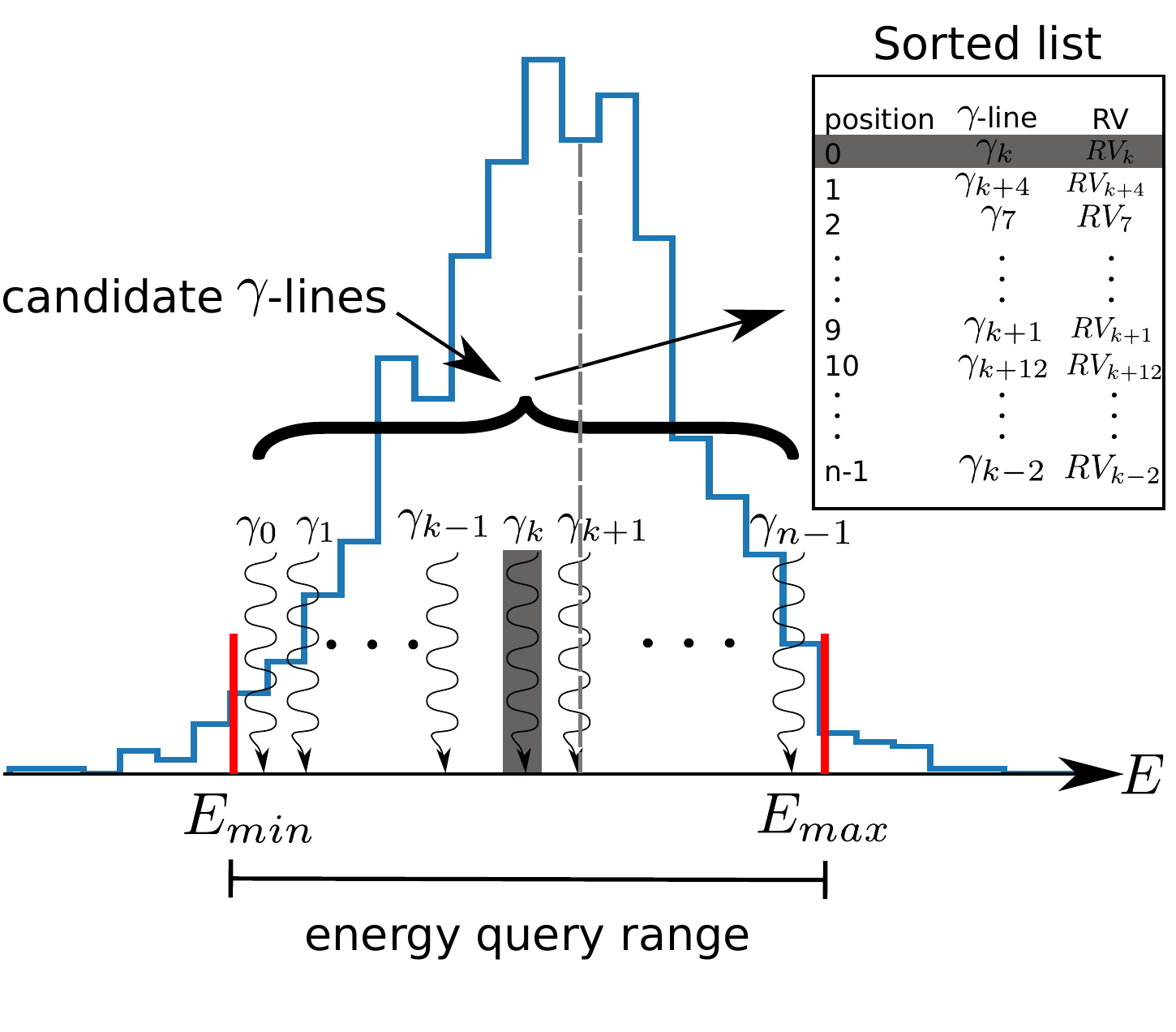}
    \caption{Graphical description of the basic process. In general, the RV is criterion dependent which normally involves other peaks. Note that the $\gamma$-line closest to the peak mean is not at the top positions because sorting depends of the RV.}
    \label{fig:coreProc}
\end{figure}

\begin{figure*}[bht]
    \centering
    \includegraphics[width=0.7\textwidth]{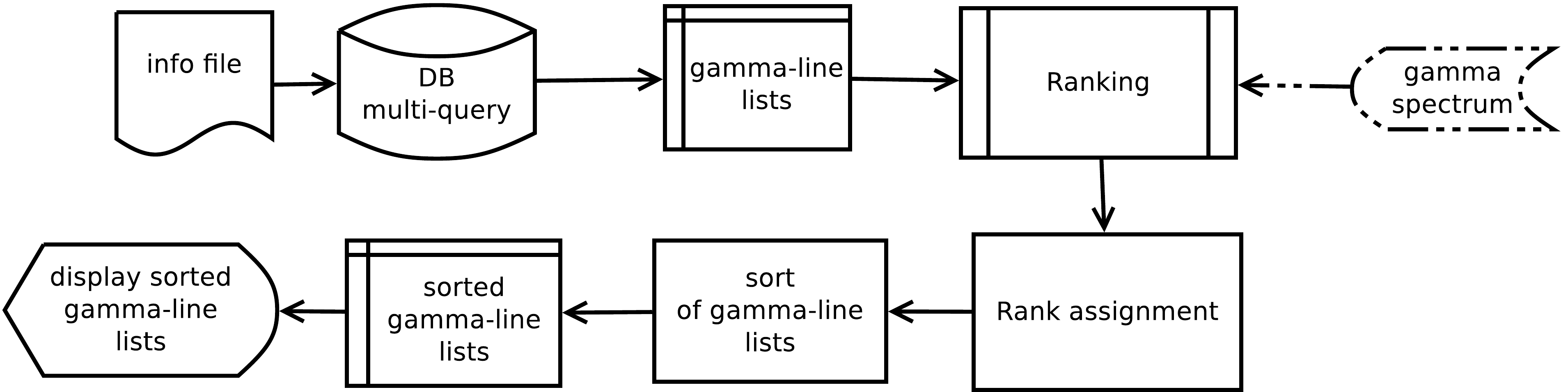}
    \caption{Flowchart of the processing done for a given info file. Some ranking
    operations may need extra information not present in the \texttt{info} 
    file such as the gamma spectrum.}
    \label{fig:flowchart2}
\end{figure*}

An \texttt{info} file \cite{histoGeUserManual} contains the energy query ranges associated with each peak. 
Once the file has been read, a 
query to the database is performed for each peak. 
As a result, a set of lists, one list per interval, are
obtained from those queries. The lists contain information 
of all $\gamma$-lines that can be located inside the
specified energy ranges. Once the RVs are calculated, each $\gamma$-line
list is sorted and printed on the
screen or stored in a text file (figure \ref{fig:coreProc}). The ranking operation calculates and assigns to each $\gamma$-line the RV used for
sorting in descending order depending on the criterion 
(section \ref{sec:crit} describes them in detail). 
This RV, or score, can be constructed so as to take into consideration global aspects of the spectrum, such as, for
example, the possibility that a given isotope or decay chain may be responsible for several peaks, bringing context into the analysis. 
Once a RV has been assigned to all the $\gamma$-line candidates under the peaks of interest, they are sorted with respect to the other candidates within the same peak (locally), positioning the best candidates at the top of the list.
Positions span from 0 to the number of the
$\gamma$-lines found in the query minus
one. A position close to 0 indicates a high preference
for the $\gamma$-line to explain features present
in the spectrum.
Although all criteria need an  \texttt{info} file, not all
criteria require the spectrum. This general procedure is schematically represented in figure \ref{fig:flowchart2}.

The ranking methods can be classified in three broad categories
according the way the RV is
calculated. Those where the RV is calculated for each
$\gamma$-line individually without considering 
the other $\gamma$-lines of the
same isotope found for the other peaks (\ref{sec:Prob}), those in which the RV is calculated for
each isotope using all the $\gamma$-lines found in
all peaks (\ref{sec:IPE},
\ref{sec:REP}, and \ref{sec:fuzzy})
and those in which the RV takes into account
all isotopes in the 
decay chains (\ref{sec:CE2}).
Depending on the category to which a criterion
belongs to, the RV is
assigned to a $\gamma$-line, or to all $\gamma$-lines
of an isotope, or to all $\gamma$-lines
of all isotopes in a chain, correspondingly.

\begin{figure}[htb]
    \centering
    \includegraphics[width=\linewidth]{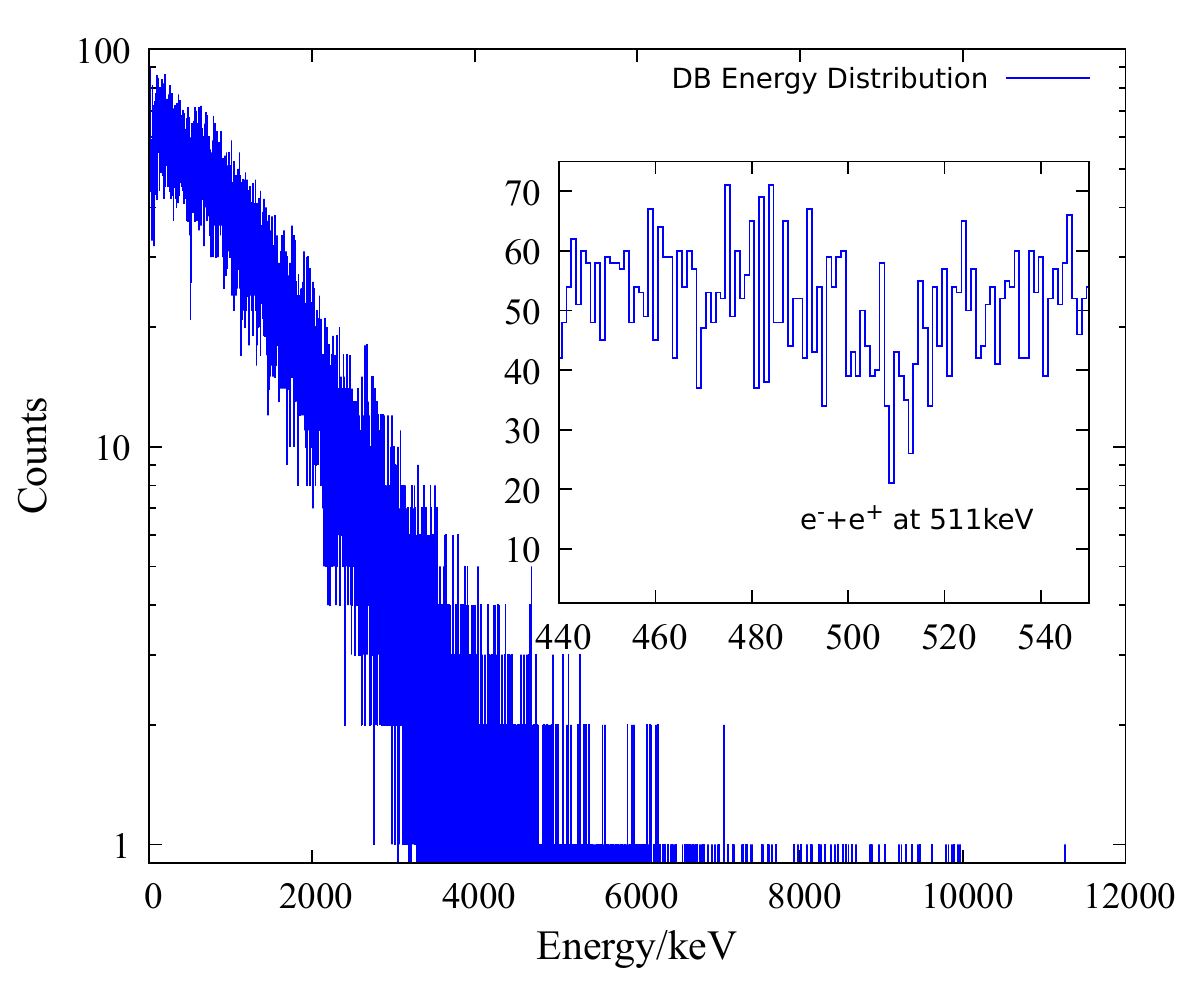}
    \caption{Energy distribution of the entries in the LDB in 1 keV bins, note the presence of the over-depleted zone (close to 511 keV) shown more clearly at the insert.}
    \label{fig:dist511}
\end{figure}

The whole operation is based on a local
database (LDB) \cite{DatabaseRepo} (11 MB of disk space) that was constructed in part
from the one accessible in \cite{nucData}. Additional data was computed and added to have enough information to be used with 
the ranking criteria such as decay chains and normalized emission probabilities \cite{histoGeUserManual}. The total number
of entries in the LDB is 92453, which is 226 times larger in comparison to 
\cite{wasim2002nucdata} and more than twice the 
used in \cite{kolotov1995software}. The database entry ($\gamma$-line) with the highest 
energy belongs to $^{20}$Na with 11258.9 keV. Figure \ref{fig:dist511} shows the energy
distribution of $\gamma$-lines in the LDB.
There is an apparent under representation in the
vicinity of 511 keV ($\approx 50$\%). This might indicate a systematic
over-subtraction of positron annihilation backgrounds in
the reported measurements.

Using a large database introduces a problem to peak identification
because peaks in a spectrum could be explained by 
many $\gamma$-lines, even if the peak has a 
narrow width. The ranking criteria presented 
in Section \ref{sec:crit} aim to overcome this issue.

\section{\label{sec:crit} Ranking Criteria}

Ten criteria were originally designed to identify 
$\gamma$-lines from potential isotopes that could be present
in a sample using the peaks found in its  $\gamma$
spectrum. They were labeled arbitrarily from A to J, however, only the 
most significant are reported here using their original names and codes.
Information about all the methods is reported in the
\texttt{histoGe}'s User Manual \cite{histoGeUserManual}. 
The criteria can be applied individually or in combinations in arbitrary order.

These criteria differ from other previously reported methods 
in how they utilize the information 
such as the emission probability (EP), the relative emission
probability (REP), the ratio 
between the number of peaks identified in the spectrum and the number
of peaks found in the database 
for certain isotopes, among others. Each criterion has merit
by itself, but they can 
be combined to obtain better results. Before the analysis begins,
the spectra must be properly calibrated to obtain reliable results.

The criteria described in this work, their code and their ranking
category are listed in table \ref{tab:Criteria}. \par

When two gamma lines in the same peak get the same RV, other criteria
are used as tiebreakers to decide in favor of one of them, e.g., for
criterion F, criterion E is used and if the tie persists, then,
criterion D is used. For criterion H, $RMSE_{mod}$ and criterion F
were used. Next, a description of each criterion is given.

\begin{table}[htb]
\caption{\centering \textsc{Criteria name, code, ranking category and the section where each is described.}}
\centering
\begin{tabular}{|c|c|c|c|}
\hline
Name                                  & Code & Ranking category & Section                        \\ \hline \hline
$\gamma$-line coincidence & \multirow{2}{*}{B}  & \multirow{2}{*}{$\gamma$-line}    & \multirow{2}{*}{\ref{sec:Prob}}  \\
probability & & &\\ \hline
Improved Peak & \multirow{2}{*}{E}  & \multirow{2}{*}{isotope}          & \multirow{2}{*}{\ref{sec:IPE}}   \\
Explanation Power & & & \\ \hline
Relative Emission  & \multirow{2}{*}{F}  & \multirow{2}{*}{isotope}          & \multirow{2}{*}{\ref{sec:REP}}   \\ 
Probability (REP) & & &\\ \hline

Fuzzy logic                           & H  & isotope          & \ref{sec:fuzzy} \\ \hline
Decay chain using REP                 & J  & chain            & \ref{sec:CE2}   \\ \hline
\end{tabular}
\label{tab:Criteria}
\end{table}

\subsection{\label{sec:Prob} $\gamma$-line coincidence probability}

The distance $d$ (in keV) from the peak mean to each of the $\gamma$-lines is used to calculate the probability
($P_G$) that the peak can be explained by a $\gamma$-line. This is done
using the cumulative distribution function (CDF) for the Gaussian
distribution:

\begin{equation}
   P_G(d,\mu=0,\sigma) = 2 \text{CDF}(-|d|,0,\sigma)
   \label{eq:certProb}
\end{equation}
\\
where $\mu$ is the mean of the CDF
and $\sigma$ is the peak's standard deviation obtained from a Gaussian fit.
The RV is assigned directly to each $\gamma$-line
and it is given by equation (\ref{eq:certProb}). The $\gamma$-lines are then sorted according to their RVs.

\subsection{\label{sec:IPE} Improved Peak Explanation Power}

A single parent isotope could have multiple $\gamma$-lines appearing as peaks in a spectrum. Every time 
a $\gamma$-line of a particular isotope is
matched with a peak, the chance 
that the isotope is present
in the sample is increased, making it more suitable to explain 
the spectrum as a whole.
The RV is calculated as the ratio of the number of $\gamma$-lines
from a given isotope that fall within the peaks of the spectrum,
to the number of $\gamma$-lines from that isotope expected in the
whole range of the spectrum (from the first to the last
peak). The RV is in the semi-closed interval (0,1] and sorting is done in
descending order.

\subsection{\label{sec:REP} Relative Emission Probability (REP)}

The REP was calculated and included in the LDB
for each $\gamma$-line. This had to be done since many of 
the entries had non-numeric or missing EP values. When not available,
EP were set to the minimum EP for the isotope \cite{histoGeUserManual}.
Using the REP allows to associate a set of $\gamma$-lines with their respective parent
isotopes knowing that, for a given isotope, the REP of its $\gamma$-lines
should add up to 1 \cite{histoGeUserManual}. 

The RV is calculated as the sum of the
REP of those $\gamma$-lines found in the queries 
for a particular isotope. 
This method also gives rank
values in the (0,1] and they are sorted in descending order.

\subsection{\label{sec:fuzzy} Ranking With a Fuzzy Inference System}

Fuzzy logic \cite{zadeh1988fuzzy} was used to compose a more powerful RV from the combination of three inputs. It is a 
method to formalize 
``approximate'' reasoning and it is a tool to treat 
uncertainty and vagueness. Unlike classical logic, where
propositions can only be true or false;
propositions in fuzzy logic can have a degree of truth between 0 and
1 \cite{ross2005fuzzy}. 
A fuzzy inference system (FIS)
performs a deductive inference through IF-THEN rules 
with fuzzy sets \cite{zadeh1965fuzzy}. A well known and straight
forward way to implement a FIS is through
the Mamdani's inference \cite{mamdani1976advances}.
The steps to perform this type of inference are:
fuzzification of the inputs; inference, which is
divided in: calculation of the antecedents, implication and
aggregation of rules; and finally, defuzzyfication via the centroid method
\cite{ross2005fuzzy} which was used to convert a fuzzy output to a crisp number, this value is used as a RV.

Three inputs (antecedents) and one output (consequent) were used
in the FIS designed to identify and rank the isotopes. The inputs of the
FIS are: a ``modified Root Mean Square Error'' ($RMSE_{Mod}$), the peak ratio
defined as $\frac{P_r}{P_T}$ where $P_r$ is the number of peaks explained
by the isotope and $P_T$ is the total
number $\gamma$-lines within
the spectrum's range, and the REP ($I_{g_R}$); the output
is a value between 0 and 1 and, in this context,
it will be called 
``Affinity''. It reflects the degree to which a set of $\gamma$-lines can be considered to belong to a particular isotope.

The $RMSE_{Mod}$ is defined using some statistical methods to analyze $\gamma$-ray spectra presented by Gilmore \cite{gilmore2011practical} such as the
net area of a peak $A = G-B$, where $G$ is the peak's integral and $B$ is the estimated background under the peak, together with the REP. Then, $RMSE_{Mod}$ is defined as:

\begin{equation}
   RMSE_{Mod} = \frac{P_{T}}{P_{r}} \sqrt{\frac{1}{N} \sum_{i=1}^{N}  \left[ \left(\frac{A_i}{A_T}\right) - \left(\frac{Ig_i}{Ig_T}  \right) \right]^2 }, 
\label{eq:MeanSquareError}
\end{equation}
\\
where $N$ is the total number
of peaks, $A_i$ is the net
area of the i-th peak, $A_T$ is the sum of
all the net areas, $Ig_i$ is the REP of the i-th peak and $Ig_T$
is the sum of the REP for all lines 
identified for the respective isotope. 
The factor $\frac{P_{T}}{P_{r}}$ that modifies the
$RMSE$ in equation (\ref{eq:MeanSquareError}), was included
to penalize those isotopes that explain fewer
peaks of the spectrum in comparison to the expected number of peaks. 

\begin{figure}[t]
\centering
\begin{subfigure}[b]{\linewidth}
\centering
\begin{tikzpicture}
\begin{axis}[xmin=-0.05,xmax=1.05,ymin=0,ymax=1,samples=1000, width=\linewidth,height=0.4\linewidth,xlabel=RMSE$_{mod}$,grid=none, title= Fuzzy sets of the FIS, legend style={draw=none,at={(0.63,0.5)}, anchor=west,  align=left}]

\addplot[blue, thick, mark=triangle,mark repeat = 20] {1/(1+exp(100*(x-0.1))))}; \addlegendentry{\tiny L:$f_s(-100,0.1)$}
\addplot[red,  thick, mark=*,mark repeat = 20, mark phase = 4] {exp((-(x-0.2)^2)/(2*0.08^2)))};
\addlegendentry{\tiny M:$f_g(0.08,0.2)$}
\addplot[green, thick, mark=x,mark repeat = 20, mark phase = 8] {1/(1+e^(-100*(x-0.3))))};
\addlegendentry{\tiny H:$f_s(100,0.3)$}
\end{axis}
\end{tikzpicture}
\end{subfigure}
\hfill
\begin{subfigure}[b]{\linewidth}
\centering
\begin{tikzpicture}
    \begin{axis}[xmin=-0.05,xmax=1.05,ymin=0,ymax=1,samples=1000, width=\linewidth,height=0.4\linewidth,xlabel=Peak Ratio,grid=none, legend style={draw=none,at={(0.025,0.5)}, anchor=west,  align=left}]

    \addplot[blue, thick, mark=triangle,mark repeat = 20, ] {1/(1+exp(50*(x-0.55))))}; \addlegendentry{\tiny L:$f_s(-50,0.55)$}
    \addplot[red,  thick, mark=*,mark repeat = 20, mark phase = 4] {exp((-(x-0.7)^2)/(2*0.125^2)))};
    \addlegendentry{\tiny M:$f_g(0.125,0.7)$}
    \addplot[green, thick, mark=x,mark repeat = 20, mark phase = 8] {1/(1+e^(-50*(x-0.85))))};
    \addlegendentry{\tiny H:$f_s(50,0.85)$}
    \end{axis}
\end{tikzpicture}
\end{subfigure}
     \hfill
\begin{subfigure}[b]{\linewidth}
\centering
 \begin{tikzpicture}
         
    \begin{axis}[xmin=-0.05,xmax=1.05,ymin=0,ymax=1,samples=1000, width=\linewidth,height=0.4\linewidth,xlabel=REP,grid=none, , legend style={draw=none, at={(0.63,0.5)}, anchor=west,  align=left}]

    \addplot[blue, thick, mark=triangle,mark repeat = 20] {1/(1+exp(50*(x-0.35))))}; \addlegendentry{\tiny L:$f_s(-50,0.35)$}
    \addplot[red,  thick, mark=*,mark repeat = 20, mark phase = 4] {exp((-(x-0.5)^2)/(2*0.125^2)))};
    \addlegendentry{\tiny M:$f_g(0.125,0.5)$}
    \addplot[green, thick, mark=x,mark repeat = 20, mark phase = 8] {1/(1+e^(-50*(x-0.65))))};
    \addlegendentry{\tiny H:$f_s(50,0.65)$}
    \end{axis} 
    \end{tikzpicture}
\end{subfigure}

\begin{subfigure}[b]{\linewidth}
\centering
\begin{tikzpicture}
    \begin{axis}[xmin=-0.05,xmax=1.05,ymin=0,ymax=1,samples=1000, width=\linewidth,height=0.5\linewidth,xlabel=Affinity,grid=none, legend style={draw=none,at={(0.6,0.5)}, anchor= west,  align=left}]

    \addplot[blue, thick, mark=triangle,mark repeat = 20] {1/(1+exp(100*(x-0.2))))}; \addlegendentry{\tiny VL:$f_s(-100,0.2)$}
    \addplot[red,  thick, mark=*,mark repeat = 20,mark phase = 4] {exp((-(x-0.3)^2)/(2*0.085^2)))};
    \addlegendentry{\tiny L:$f_g(0.085,0.3)$}
    \addplot[green, thick, mark=x,mark repeat = 20, mark phase = 8] {exp((-(x-0.5)^2)/(2*0.085^2)))};
    \addlegendentry{\tiny M:$f_s(0.085,0.5)$}
    \addplot[cyan,  thick, mark=square,mark repeat = 20, mark phase = 12] {exp((-(x-0.7)^2)/(2*0.085^2)))};
    \addlegendentry{\tiny H:$f_g(0.085,0.7)$}
    \addplot[black, thick, mark=circle*,mark repeat = 20, mark phase = 16] {1/(1+e^(-100*(x-0.8))))};
    \addlegendentry{\tiny VH:$f_s(100,0.8)$}
\end{axis}
\end{tikzpicture}
\end{subfigure}
\caption{Parameters of the fuzzy sets used in the FIS. Sigmoid and Gaussian functions parameters are $f_s(x,k_o,x_o)$ and $f_g(x,\sigma,\mu)$, respectively. The names of the fuzzy sets are: VL is ``Very Low'', L is ``Low'', M is ``Medium'', H is ``High'' and VH is ``Very High''.}
\label{fig:fuzzySet}
\end{figure}
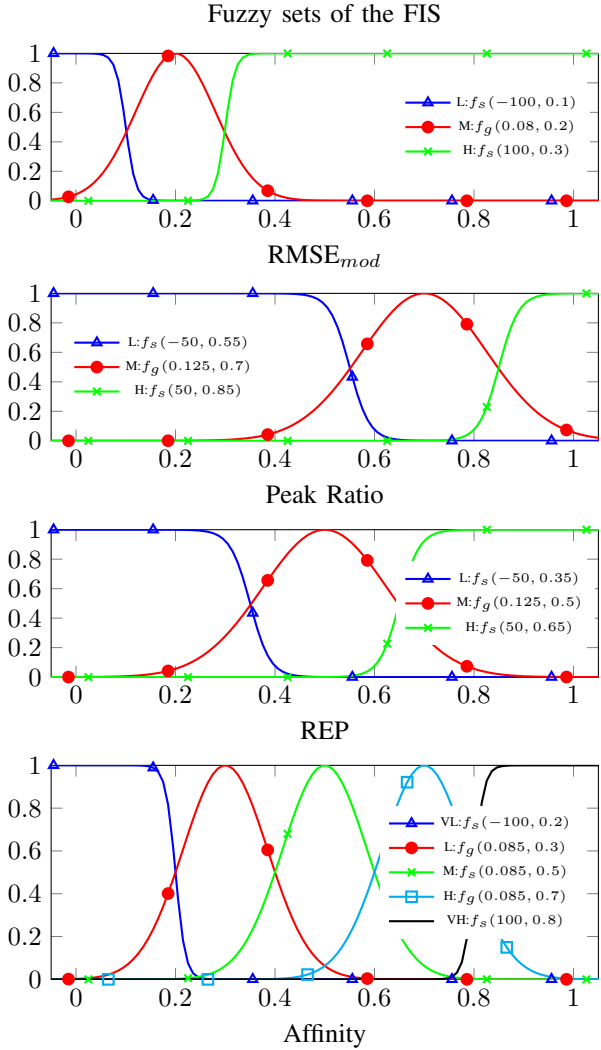

Each input 
has three fuzzy sets and the output has five fuzzy
sets. Fuzzy sets were defined using the well-known sigmoid and Gaussian functions
whose mathematical expressions are,
respectively, given by:

\begin{equation}
   f_s(x,k_o,x_o) = \frac{1}{1 + \large{e}^{-k_o(x-x_o)}},
\label{eq:SigmoidSet}
\end{equation}
\\
where $k_o$ and $x_o$ are the parameters of the sigmoid function, and 

\begin{equation}
  f_g(x,\sigma,\mu) = \large{e}^ \frac{-(x-\mu)^2}{2\sigma^2}, 
\label{eq:GaussianSet}
\end{equation}
\\
where $\sigma$ and $\mu$ are the parameters of the
Gaussian function. Figure \ref{fig:fuzzySet} shows the name, curve, the mathematical function of the fuzzy sets used in the FIS.
The parameters of the fuzzy
sets were established considering the designer's own knowledge about what the
linguistic variables Very Low (VL), Low (L),
Medium (M), High (H) or Very High (VH) could mean considering that the output hypersurface must be a monotonically
increasing one.

The rules used for the inference are of the form: 

\begin{center}
    
IF $x_1$ is $RMSE_m^k$ and $x_2$ is $\textrm{PeakRatio}_n^k$ and  $x_3$ is $\textrm{Ig}_{R_p}^k$ THEN $y^k$ is $\textrm{Affinity}_q^k$,
\end{center} 
where $k$ is an index that refers to $k$-th rule as is shown in table \ref{tab:FuzzyRules} and $m$, $n$, $p$ and $q$ are
the indices of their respective fuzzy sets which can be VL, L, M, H or VH
depending on whether they are input or output fuzzy sets, as shown in figure \ref{fig:fuzzySet}. The
rules used for the inference process are shown in table
\ref{tab:FuzzyRules}. Since there are three
fuzzy sets for each input, there are twenty seven rules
with their consequents chosen from five fuzzy sets. 
During the design of the FIS, it was decided that
five output fuzzy sets were enough to categorize the combinations of the inputs. \par

Once the fuzzy sets are defined and the rules given,
the Mamdani inference can be implemented to calculate the affinities of isotopes.
All $\gamma$-lines that share the same parent get the same RV.
They are later sorted in descending order using as a tiebreaker 
criterion the $RMSE_{Mod}$. This criterion was chosen for its simplicity
and because it is already an input of the fuzzy rank method,
however, other criteria such as $Ig_R$ could be used to untie
isotopes with the same RV. Efficiency corrected spectra were used,
but no important improvements were observed in the results got with this method.  

\begin{table*}[!ht]
\centering
\caption{\centering \textsc{FIS rules used in \texttt{histoGe}'s fuzzy rank. MVL, ML, MM, MH and MVH refer to the affinity output fuzzy sets: Very Low, Low, Medium, High and Very High, respectively.}}
\label{tab:FuzzyRules}
\resizebox{\linewidth}{!}{
\begin{tabular}{|c|c|c|c||c|c|c||c|c|c|} 
\hline 
& \multicolumn{3}{c|}{$RMSE_{\mathrm{Low}}$}  &  \multicolumn{3}{c|}{$RMSE_{\mathrm{Medium}}$} &  \multicolumn{3}{c|}{$RMSE_{\mathrm{High}}$} \\
\hline \hline
 & \footnotesize{$\textrm{PeakRatio}_{\mathrm{Low}}$} & \footnotesize{$\textrm{PeakRatio}_{\mathrm{Medium}}$} & \footnotesize{$\textrm{PeakRatio}_{\mathrm{High}}$} & \footnotesize{$\textrm{PeakRatio}_{\mathrm{Low}}$} & \footnotesize{$\textrm{PeakRatio}_{\mathrm{Medium}}$} & \footnotesize{$\textrm{PeakRatio}_{\mathrm{High}}$} & \footnotesize{$\textrm{PeakRatio}_{\mathrm{Low}}$} & \footnotesize{$\textrm{PeakRatio}_{\mathrm{Medium}}$} & \footnotesize{$\textrm{PeakRatio}_{\mathrm{High}}$}\\
\hline
 \footnotesize{$\textrm{REP}_{ \mathrm{Low}}$} & ML & MM & MM & MVL & ML & ML & MVL & MVL & MVL \\ \hline
 \footnotesize{$\textrm{REP}_{\mathrm{Medium}}$} & MM & MH & MH & MM & MM & MH & ML & MM & MH \\ \hline
 \footnotesize{$\textrm{REP}_{\mathrm{High}}$} & MVH & MVH & MVH & MH & MH & MVH & MH & MH & MH \\ \hline
 \end{tabular}}

\end{table*}

\subsection{\label{sec:CE2} Chain using Relative Emission Probability}

In some spectra, the presence of some peaks can be due to 
$\gamma$-lines from several 
isotopes that are connected to each other 
via a decay chain.  This motivates the design of a criterion that rank chains instead
of isotopes alone.

For this criterion, the RV is calculated as follows: REP of the $\gamma$-lines
found in the query ranges belonging to all the isotopes in a given chain are
summed and averaged. As described in section \ref{sec:genProc}, the RV is assigned
to all the $\gamma$-lines of all isotopes that belong to the chain.
Sorting is done in descending order.

\section{\label{sec:experimental} Experimental Setup}
The results in sections \ref{sec:rsproc} and \ref{sec:results} below were obtained with spectra from a set of radioactive sealed 
calibration sources (Table \ref{tab:sources}) and they were acquired with an EG\&G-ORTEC Hyper Pure Germanium detector in 
the Detectors Laboratory at the Institute of Nuclear Science of the 
National Autonomous University of Mexico (UNAM), hereafter referred to
as ICN-HPGe detector, whose characterization has been reported elsewhere 
\cite{Aguilar2020GeDetector}. The data acquisition system (DAQ) was a PX5-HPGe
multi-channel analyzer (MCA) and a digital pulse processor (DPP)
software analyzer provided by Amptek \cite{amptek}. 

\begin{table}[!b]
\centering
\caption{\centering \textsc{\label{tab:sources} Energy and half-life of the point-like sealed sources used to compare ranking criteria. All sources had an initial activity of 1}~$\mu$Ci \textsc{(except $^{137}$}Cs \textsc{with} 0.1~$\mu$Ci \textsc{, as of January 2019 and a 20\% uncertainty,} \textsc{and } $^{241}$Am \textsc{with} 121nCi \textsc{as of August 1982}).}

\begin{tabular}{|c|c|c|}
\hline
 Source & $\gamma$-lines [keV]  & half-life [yr]\\
 \hline
 \multirow{2}{*}{$^{241}$Am} &  13.81, 27.03, 33.19, 43.42, 59.54, 69.76 & \multirow{2}{*}{432.5} \\ & 98.97, 102.98, 120.36, 125.33 &   \\ \hline
 \multirow{2}{*}{$^{133}$Ba} & 53.1, 79.6, 81.0, 160.6, 223.3, 276.3 & \multirow{2}{*}{10.5} \\ & 302.8, 356.0, 383.8 & \\ \hline
 $^{109}$Cd & 88.07 & 1.27 \\ \hline
 $^{57}$Co & 122.0, 136.0& 0.745 \\ \hline
 $^{60}$Co & 1173.2, 1332.5 & 5.27 \\
 \hline
 $^{137}$Cs & 662.0 & 30.1 \\
 \hline
 $^{54}$Mn & 835.0 & 0.855 \\
 \hline
 $^{22}$Na & 1275.0 & 2.6 \\ \hline
 $^{65}$Zn & 1115.0 & 0.668 \\\hline

\end{tabular}
\end{table}

\section{\label{sec:results_complete} Results}
\subsection{\label{sec:rsproc}Example with a $^{60}$Co sealed point-like source}
\begin{figure*}
    \centering
    \includegraphics[scale=0.33]{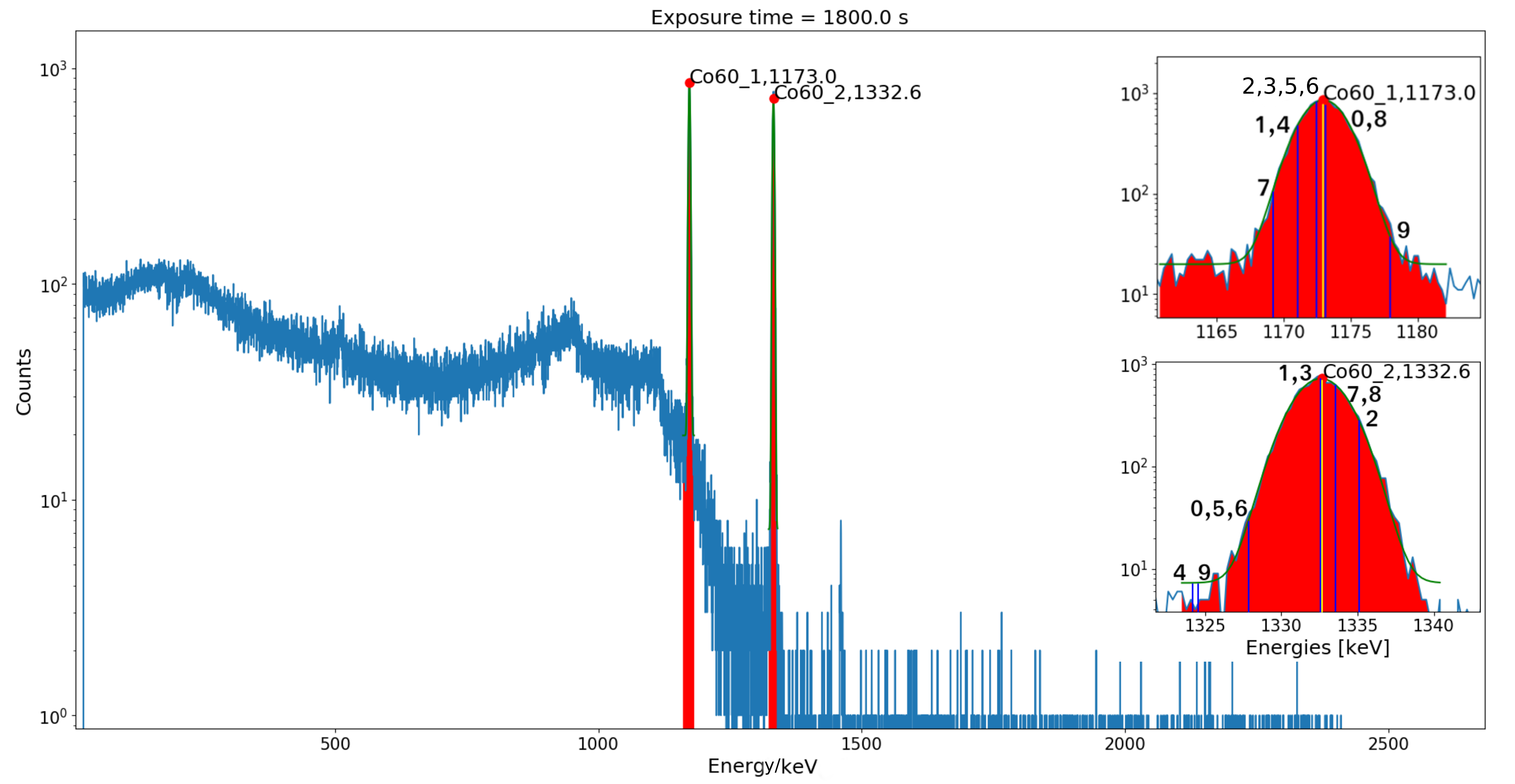}
    
    \caption{$^{60}$Co spectrum obtained from the point-like source described in table \ref{tab:sources}. Peak ranges for
    energy queries are drawn in red. Plot was made using \texttt{histoGe} \cite{histoGeRepo}.
    The zoomed-in inserts show the peaks along with first 10 $\gamma$-lines
    (tagged by their position) in accordance to table \ref{tab:r360CoSource1}.} 
    \label{fig:60Co}
\end{figure*}

As a first example, criterion F (\ref{sec:REP})
was applied over the spectrum obtained in the 
ICN-UNAM from the $^{60}$Co point-like source reported in table 
\ref{tab:sources}. For this test, an 
\texttt{info} file with two query ranges,
associated with each of the two more prominent
gamma lines of this isotope was used. Table 
\ref{tab:r360CoSource1} shows the isotopes
in the top 10 positions output and their corresponding gamma lines.
The two energy ranges
are shown in red in
figure \ref{fig:60Co} as well as the 10 best
candidates $\gamma$-lines are shown in the inserts.  

Notice that, since this criterion assigns a RV per isotope,
$\gamma$-lines from the same isotope found in the two query
ranges have the same RV. For the first query range (Co60\_1:
from 1160.84 keV to 1182.18 keV), the isotope $^{60}$Co is
found at the top position by itself, but for the other one
(Co60\_2: 1323.41 keV to 1340.54 keV) it is tied with
$^{53}$Co, since all expected gamma lines are found in the
query ranges for both isotopes. A tie breaking based on
criterion E is effected, which places $^{60}$Co at the top ($^{\dagger}$).

\begin{table}[ht]
\centering
\caption{\centering \textsc{\label{tab:r360CoSource1} The energy, gamma intensity, parent isotope and rank value for the top 10 isotopes found in each of the 2 query ranges (}Co60\_1, Co60\_2\textsc{), obtained from running criterion F on the $^{60}$}Co \textsc{sealed source spectrum. See text for details.}}

\begin{tabular}{|c|c|c|c|c|}
\hline
\multicolumn{5}{|c|}{\text{$\text{Co60}_1$: from 1160.84 to 1182.18 PBR = 8.2 }}\\
\hline
 Eg [keV]  & Ig (\%)  & Parent & Rank Value & Position  \\
 \hline \hline
\rowcolor[gray]{.9}   1173.237 (4) &   99.97 &    $^{60}$Co & 1.000 & 0\\ \hline
   1171.3 (2) &    1.70  &   $^{120}$Sb & 0.889 & 1\\ \hline
   1172.9 (1) &   98.00  &   $^{62m}$Co & 0.806 & 2\\ \hline
   1172.9 (1) &   98.00  &   $^{62m}$Co & 0.806 & 3\\ \hline
   1171.3 (2) &   19.00  &   $^{120}$In & 0.731 & 4\\ \hline
   1172.9 (1)&    84.00  &    $^{62}$Co & 0.722 & 5\\ \hline
   1172.9 (1) &   0.34   &    $^{62}$Cu & 0.579 & 6\\ \hline
   1168.8 (5) & 100.00   &  $^{128m}$Sn & 0.491 & 7\\ \hline
 1173.237 (4) &   0.26   &    $^{60}$Cu & 0.442 & 8\\ \hline
    1178.5 () &  64.00   &    $^{34}$Si & 0.400 & 9\\ \hline \hline
\multicolumn{5}{|c|}{\text{$\text{Co60}_2$: from 1323.41 to 1340.54 PBR = 24.4}}\\
\hline
    
    Eg [keV]  & Ig (\%) & Parent & Rank Value & Position \\
 \hline \hline
\rowcolor[gray]{.9}  1332.501 (5) &  99.99 & $^{60}$Co  &  1.000 & 0$^\dagger$ \\ \hline
 1328.2 (3) &   5.60 & $^{53}$Co  &  1.000 & 1\\ \hline
 1335.04 (10) &  71.00 & $^{125}$In &  0.693 & 2\\ \hline
 1332.501 (5) &  88.00 & $^{60}$Cu  &  0.442 & 3\\ \hline
   1324.1 (2) &   0.47 & $^{210}$At &  0.347 & 4\\ \hline
   1328.2 (3) &  86.00 & $^{53m}$Co &  0.305 & 5\\ \hline
   1328.2 (3) &  87.00 & $^{53m}$Fe &  0.304 & 6\\ \hline
   1333.4 (3) &   0.52 & $^{142}$Cs &  0.221 & 7\\ \hline
   1333.4 (3) &   0.52 & $^{142}$Cs &  0.221 & 8\\ \hline
  1324.51 (6) &  17.50 & $^{150}$Pm &  0.192 & 9\\ \hline

\end{tabular}

\end{table}

\subsection{\label{sec:results} Comparison of criteria using known $\gamma$-sources}

To compare the performance of the criteria discussed in section \ref{sec:crit}, 
\texttt{histoGe} was employed to identify the $\gamma$-lines from point-like sealed sources of table \ref{tab:sources} through their $\gamma$ spectra. 
Table \ref{tab:AllIsotopes} summarizes the results, showing the positions after sorting the
RV of their corresponding $\gamma$-lines, query ranges and $\gamma$-line properties. The positions given by \texttt{InterSpec} \cite{interspec} are shown for comparison. \par

\begin{table*}[th]
\centering
\caption{\centering \textsc{\label{tab:AllIsotopes} Position for every
$\gamma$-line expected from  the spectra of the radioactive sources of table \ref{tab:sources} using the
criteria described in section \ref{sec:crit}. The isotopes and
their $\gamma$-lines properties, energy query range, the
number of isotopes found in the query range, the positions of each $\gamma$-line are shown for each criterion and,
besides, the IS's individual performance.
The top $\gamma$-line position is zero. $^{210}$}Pb \textsc{spectra was obtained from sample A from
table \ref{tab:waterUk}. For} $^{241}$Am \textsc{ranges could include more than one $\gamma$-line reported in the
database, in this case all are ranked together.}}
\scalebox{1.0}{
\begin{tabular}{|c|c|c|c|c|c|c|c|c|c|c|c|c|c|}\hline
& \multicolumn{2}{|c|}{$\gamma$-line Properties} &  \multicolumn{2}{|c|}{Query Range} & \multicolumn{2}{|c|}{} &\multicolumn{7}{|c|}{$\gamma$-line Position} \\\hline
                            & Eg [keV]  &  Ig (\%)& Emin & Emax & PBR &Size &B & E & F & H & H+ & J & IS\\ \hline \hline
\multirow{13}{*}{$^{241}$Am}& 26.344 & 2.4  & 24.187  & 28.355   & 10.5 & 200 & 0   & 75  & 7 & 1 & 1 & 34 & 4  \\
                            & 33.196 & 0.13 & 31.122  & 35.550   & 1.7  & 198 & 0   & 61  & 3 & 1 & 0 & 41 & 1   \\
                            & 43.423 & 0.07 & 42.355  & 44.483   & 0.2  & 143 & 1   & 50  & 5 & 5 & 0 & 29 & 9   \\
                            & 51.010 & 0.00 & 48.031  & 51.869   & 0.2  & 194 & 158 & 45  & 2 & 1 & 0 & 33 & 9   \\ 
                            & 59.541 & 35.9 & 56.673  & 63.802   & 19.5 & 414 & 0   & 114 & 9 & 3 & 1 & 82 & 1   \\
                            & 69.760 & 0.00 & 68.316  & 71.735   & 0.03  & 206 & 28  & 50  & 3 & 0 & 0 & 42 & 1   \\
                            & 75.800 & 0.00 & 75.411  & 78.530   & 0.1  & 177 & 153 & 53  & 5 & 4 & 2 & 39 & 9   \\
                            & 79.100 & 0.00 & 78.649  & 82.377   & 0.1  & 211 & 163 & 56  & 5 & 2 & 0 & 35 & 13   \\
                            & 98.970 & 0.02 & 96.695  & 100.024  & 0.5  & 251 & 17  & 81  & 1 & 5 & 0 & 44 & 4   \\
                            & 102.98 & 0.02 & 100.502 & 105.081  & 0.5  & 345 & 2   & 104 & 3 & 2 & 0 & 51 & 5   \\
                            & 120.36 & 0.00 & 116.561 & 121.999  & 2.9  & 372 & 198 & 124 & 4 & 1 & 1 & 24 & 4   \\
                            & 125.30 & 0.00 & 123.656 & 127.075  & 0.6  & 239 & 1   & 74  & 5 & 0 & 0 & 38 & 2   \\ \hline \hline
\multirow{9}{*}{$^{133}$Ba} &53.161  & 2.2   & 51.353  & 56.238  & 8.3   & 258 & 84  & 0 & 0 & 0 & 0 & 5  &n/a\\
                            &79.613  & 2.62  & 78.327  & 84.538  & 5.1   & 344 & 206 & 1 & 1 & 0 & 0 & 8  &3\\
                            &80.997  & 34.06 & 78.327  & 84.538  &     -   & 344 & 206 & 1 & 1 & 0 & 0 & 8  &6\\
                            &160.613 & 0.65  & 156     & 166     & 0.06   & 681 & 61  & 0 & 0 & 1 & 1 & 9  &0\\
                            &223.234 & 0.45  & 220     & 227     & 0.06   & 220 & 52  & 0 & 0 & 0 & 0 & 17 &0\\
                            &276.398 & 7.16  & 273.776 & 279.103 & 7.5   & 273 & 41  & 0 & 0 & 0 & 1 & 10 &2\\
                            &302.853 & 18.33 & 299.866 & 306.077 & 7.5   & 347 & 10  & 0 & 0 & 0 & 0 & 10 &11\\
                            &356.017 & 62.05 & 352.929 & 360.140 & 197.6  & 382 & 11  & 0 & 0 & 0 & 0 & 4  &0\\
                            &383.851 & 8.94  & 380.345 & 387.556 & 29.5  & 453 & 7   & 0 & 0 & 0 & 0 & 14 &1\\ 
                            \hline \hline

$^{109}$Cd                  &88.04(5) & 3.61 & 85.439 & 91.561 &  9.7 & 378 & 5 & 2 & 2 & 4 & 4 & 1 &1\\\hline \hline

\multirow{2}{*}{$^{57}$Co}  & 122.0614     & 85.6  & 120.080  & 124.737 & 11.4 & 315 & 70 & 2 & 3 & 0 & 0 & 0 &1\\
                            & 136.474      & 10.68 & 135.534  & 138.000 & 17.2 & 135 & 52 & 2 & 2 & 0 & 0 & 0 &2\\\hline \hline

\multirow{2}{*}{$^{60}$Co} & 1173.237 & 99.97 & 1170.839 & 1176.184 & 8.2  & 222 & 28 & 0 & 0 & 0 & 0 & 0 & 0\\
                            &1332.501 & 99.99 & 1329.410 & 1336.537 & 24.4 & 89  & 2  & 0 & 0 & 0 & 0 & 0 & 0\\\hline \hline

$^{137}$Cs & 661.657  & 85.1  & 660.302   & 663.393  & 13.0 &142 & 2  & 1 & 1 & 1 & 1 & 0 & 0\\\hline \hline

$^{54}$Mn & 834.848  & 99.98 & 831.232  & 838.346  & 19.1 & 343 & 61 & 0 & 0 & 0 & 0 & 251 & 0\\\hline \hline

$^{22}$Na & 1274.53  & 99.94 & 1270.843 & 1277.957 & 22.2 &253 & 14 & 0 & 0 & 0 & 0 & 0 & 0\\\hline \hline

$^{210}$Pb & 46.539   & 4.25 & 45.704   & 47.067   & 0.9 &68   &31 & 1 & 1 & 0 & 0 & 1 & 1\\\hline \hline

$^{65}$Zn & 1115.546 & 50.6  & 1101.568 & 1133.256  & 3.3 & 1151 & 12 & 5 & 1 & 4 & 1 & 4 & 0\\\hline 
 
\end{tabular}}
\end{table*}

\pgfplotsset{compat=1.17,every axis/.append style={font=\small,tick style={thin}}}
\usetikzlibrary{patterns}

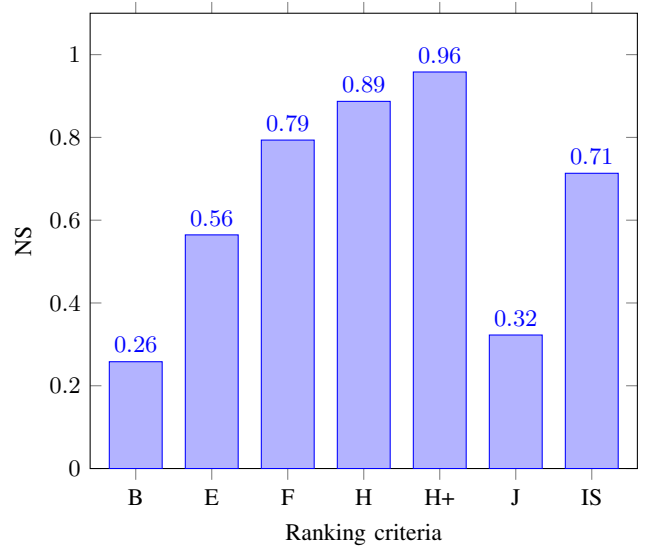
\begin{figure}[htb]
    \centering
    \begin{tikzpicture}

\begin{axis} [
    grid=none,
    xtick=data,
    symbolic x coords={B,E, F, H, H+, J, IS},
    nodes near coords,
    ybar = 0.2cm, 
    bar width = 20pt,
    ylabel= NS,
    xlabel= Ranking criteria,
    ymin = 0,
    ymax = 1,
    enlarge x limits = {.10},
    enlarge y limits = {value = .10, upper},
    legend pos=north west,
]
    
\addplot coordinates {(B,0.25806451612903225) (E, 0.5645161290322581)
(F, 0.7935483870967742) (H, 0.8870967741935484) (H+, 0.9580645161290322)
(J, 0.3225806451612903) (IS, 0.7133333333333334)};

\end{axis}
\end{tikzpicture}
\caption{Comparison of the performance of the criteria in identifying the $\gamma$-lines present in the point-like sources of Table \ref{tab:sources} and an extended $^{210}$Pb source through the normalized score (NS) defined by (\ref{eq:normalizedScore}). The score of InterSpec software \cite{interspec} is shown in the last bar (IS).}
\label{fig:scores}
\end{figure}

Figure \ref{fig:scores} compares the
performance of the various criteria using an \textit{ad hoc} metric which is
calculated as follows: a score (S) is assigned
to each $\gamma$-line (table \ref{tab:AllIsotopes}),
in the following way: 

\begin{equation}
    \mathrm{S_{criterion}}(p_{\gamma-line}) = \begin{cases} 
          10 - p_{\gamma-line} & p_{\gamma-line} < 10  \\
          0 & p_{\gamma-line} \geq 10 
       \end{cases},
\end{equation}
\\
where $p_{\gamma-line}$ is the list position of 
the $\gamma$-line (starting from 0) for a
given criterion. For example,
in the $^{109}$Cd radioactive source
(table \ref{tab:AllIsotopes}), the
$\gamma$-line located at $88.04$ keV
was sorted by criterion F at position 2 
(third place) an it receives a
score of $10-2=8$.

Then, the following operation is done 
for obtaining the normalized score (NS)
for a specific criterion:

\begin{equation}
    \mathrm{NS_{criterion}} = \frac{\sum_{\gamma-line}\mathrm{S_{criterion}(p_{\gamma-line})}}{\mathrm{MS}}
    \label{eq:normalizedScore}
\end{equation}
\\
where the maximum possible score is given by MS =
$10\times \mathrm{Total_{\gamma-lines}}=10 \times 27 = 270$. 
The NS range is between 0 and 1, being 0 and 1 the worst and the best possible results, respectively. 

Under these expressions, criterion B
performed poorly (the expected $\gamma$-line positions are mostly above 9),
it was taken as the baseline criteria
during comparison with other methods. This means that 
coincidence probability is not relevant because it depends on a good calibration and its 
uncertainty. The fact that the density of $\gamma$-lines is so high affects the 
performance, because many other $\gamma$-lines could be as near as or nearer than that of interest. 
Criterion
E was capable to identify $^{133}$Ba and
$^{60}$Co but showed a poor performance for
$^{241}$Am because it has 171 $\gamma$-lines in
\cite{DatabaseRepo} and 132 $\gamma$-lines in the range of its spectrum but only 22
$\gamma$-lines were observed. Criterion F
improves the results obtained with the
criterion E. In the worst case, it equals the performance of
criterion E, but the fact that it uses the REP makes it able to focus on those
$\gamma$-lines that have the highest EP giving
little importance to those $\gamma$-lines that
could be undetectable with a certain detector.
Criterion H gave the best results, in particular, when  those isotopes with a relatively small REP are filtered to discard them from sorting and ranking (H+). Fuzzy ranking is able to get approximately 0.95 or 0.87
of the maximum score with and without filtering (H and H+), respectively. 
The overall performance of H+ makes it
the best one with a score of 0.955. This result suggests that combining the criteria to make identification algorithms can give better results than using individual criteria alone.    
For criterion J, fair results
were obtained. A simple inspection of the results reveals that this criterion correctly identifies the $\gamma$-lines 7 of the 10 radioactive sources in Table \ref{tab:Criteria}. In particular, the $\gamma$-lines for $^{241}$Am all get ranked at positions higher than 34 and the line of $^{54}$Mn was placed at position 251. In general, $^{241}$Am was hard to be identified by \texttt{histoGe} and this penalizes some criteria more than others.

Nonetheless, no significant improvements were observed when
efficiency corrected spectra were used as input.
Therefore, identification could be achieved without knowing
the detector's efficiency \cite{Aguilar2020GeDetector}. Besides,
a test was implemented to find the dependence of the ranking
results against the peak-to-background ratio (PBR)
defined as the ratio of the area enclosed by a peak to the area of the
background beneath it, in an interval $\pm 3 \sigma$ around the peak,
and calculated using an exponential plus second order polynomial fit.
Fake peaks with variable amplitude were introduced
in known spectra and it was found that, once the PDA detect the peak,
the position of that $\gamma$-line is unaffected for methods
E, F and J, and negligible changes in position were observed for
B, H and H+. Thus, identification of peaks with low PBR are not affected.

\par

\subsection{\label{sec:crosscheck} Performance over generic $\gamma$ spectra}

To explore the capability of the program to identify
isotopes in spectra that may contain an arbitrary
combination of isotopes, various test were performed using 
previously studied samples \cite{Aguilar2020GeWaterReduction,Aguilar2020flux}: one had traces of $^{210}$Pb
(sample A) and the other contained $^{177}$Lu and $^{131}$I
(sample B) and a rock sample taken from the inside of LabChico's
site (sample C). It was known that sample A was
spiked with lead and it was acquired with the high
purity germanium detector of the Institute of Physics 
of the UNAM (IF-BEGe detector) 
\cite{Aguilar2020GeDetector} with a exposure time of 24 h,
spectrum of sample B was acquired inside the Boulby Underground Germanium Suite (BUGS) facility
\cite{SCOVELL2018160} at Boulby Underground Laboratory \cite{Malczewski2013} and a previous analysis of sample B
identified $^{177}$Lu and $^{131}$I \cite{Aguilar2020GeWaterReduction} and sample C is a rock taken during geotechnical studies \cite{Aguilar2020flux} and assayed with the ICN-UNAM HPGe \cite{Aguilar2020GeDetector}. 

Table \ref{tab:waterUk} shows
the results of applying criteria F and H to samples
A, B and C. Criteria F and H were chosen for this
test because they showed the best performance in the test with the point-like sealed sources.
For the sample A, in addition to the $^{210}$Pb $\gamma$-line at 46.5 keV  
(positions 1 for rank F and 0 for rank H), two X-rays associated with $^{210}$Pb were observed, however, their assigned positions (higher than 20) are irrelevant because the database does not contain reliable information about X-rays. This motivates the addition of X-ray information to the LDB. For sample B, ranks H and H+ identified
$^{131}$I, $^{177}$Lu and $^{40}$K at positions 0 for all
the $\gamma$-lines which is the best possible result. 
The $^{40}$K seen in sample
B can be safely attributed to the water given the 
ultra low background of the BUGS detector
\cite{scovell2018low}. About sample C, $^{40}$K was
successfully identified by ranks F, H and H+. The 
photopeak observed at 2614 keV was attributed to 
$^{208}{\rm Tl}$ because it belongs to $^{232}$Th decay 
chain and was found at positions 3 for ranks F, H and H+.
Two photopeaks were attributed to $^{214}$Bi because they belong
to $^{238}$U decay chain, however, 
no further peaks that could have increased the confidence were
identified due to the high background at energies below the 
potassium peak in the spectrum. $^{214}$Bi has 214 entries
at the database but only two were considered resulting in a low 
REP which affected the behavior of H rank.

\begin{table*}[!ht]
\centering
\caption{\centering \textsc{\label{tab:waterUk} Positions and ranks, according to criteria F, H and H+, of parent isotopes known to be present in the two water samples (A and B), and at the intended site for LABChico (C). The query ranges and $\gamma$-ray properties are also shown. *$^{214}$Bi \text{has 214 $\gamma$-line and due to only 2 lines were identified, REP is small which affects the behavior of H rank.}}}
\begin{tabular}{|c|c|c|c|c|c|c|c|c|c|c|c|c|c|}\hline
\multicolumn{2}{|c|}{} & \multicolumn{2}{|c|}{$\gamma$-ray properties}  & \multicolumn{2}{|c|}{Query Range}& & &  \multicolumn{3}{|c|}{Position} &  \multicolumn{3}{|c|}{RV} \\
\hline
 sample             & isotope    & Eg [keV] &  Ig (\%) & Emin    & Emax    & PBR        & Size & F & H  & H+ &  F    & H     & H+    \\ \hline \hline
 A                  & $^{210}$Pb & 46.53    &   4.25   & 45.70   & 47.06   & 0.950      & 69   & 1 & 0  & 0  &  1    & 0.873 & 0.873 \\ \hline
\multirow{6}{*}{B}  & $^{177}$Lu & 112.94   & 6.4      & 111.11  & 115.14  & 1.010      & 279  & 1 & 0  & 0  & 0.965 & 0.904 & 0.999 \\
                    & $^{177}$Lu & 208.36   & 11.0     & 206.8   & 209.8   & 2.489      & 191  & 3 & 0  & 0  & 0.965 & 0.904 & 0.999 \\
                    & $^{131}$I  & 80.18    & 2.62     & 79.3    & 82.0    & 0.169      & 158  & 4 & 0  & 0  & 0.969 & 0.912 & 0.999 \\
                    & $^{131}$I  & 284.30   & 6.14     & 282.5   & 285.5   & 2.289      & 176  & 0 & 0  & 0  & 0.969 & 0.912 & 0.999 \\
                    & $^{131}$I  & 364.48   & 81.7     & 362.51  & 366.1   & 11.321     & 189  & 0 & 0  & 0  & 0.969 & 0.912 & 0.999 \\
                    & $^{131}$I  & 636.98   & 7.17     & 634.64  & 639.0   & 3.550      & 237  & 0 & 0  & 0  & 0.969 & 0.912 & 0.999 \\
                    & $^{40}$K   & 1460.83  & 11.0     & 1457.0  & 1463.53 & 541.346    & 158  & 0 & 0  & 0  & 1.000 & 0.873 & 0.873 \\ \hline
 \multirow{3}{*}{C} & $^{40}$K   & 1460.83  & 11.0     & 1450.07 & 1467.09 & 0.436      & 423  & 0 & 0  & 0  & 1.000 & 0.873 & 0.873 \\
                    & $^{214}$Bi*& 1693.3   & 0.01     & 1690.0  & 1694.0  & 0.106      & 88   & 1 & 19 & 0  & 0.117 & 0.120 & 0.468 \\
                    & $^{214}$Bi*& 1764.49  & 15.4     & 1762.45 & 1766.96 & 0.383      & 92   & 3 & 19 & 1  & 0.117 & 0.120 & 0.468 \\
                    & $^{208}$Tl & 2614.5   & 99.0     & 2600    & 2620    & 0.957      & 150  & 3 & 3  & 3  & 0.432 & 0.417 & 0.674 \\\hline
\end{tabular}
\end{table*}

\begin{table*}[!ht]
\centering
\caption{\centering \textsc{\label{tab:gSasso} RV and positions calculated for 4 criteria from the published parent isotopes on figure 2 of \cite{granSasso}. Note from the size column that the number of candidates is in the hundreds.}}
\begin{tabular}{|c|c|c|c|c|c|c|c|c|c|c|c|c|c|c|} 
\hline
\multicolumn{1}{|c|}{} & \multicolumn{2}{|c|}{$\gamma$-line properties} &\multicolumn{2}{|c|}{Query Range [keV]}&  & \multicolumn{2}{|c|}{Position}  &    \multicolumn{2}{|c|}{RV} \\
\hline
 isotope    & Eg [keV] & Ig (\%) & Emin   & Emax   & Size& F & J   & F     & J         \\ \hline
 $^{210}$Pb & 46.53    & 4.25    & 45.42  & 49.98  & 227 & 3 & 4   & 1     & 0.288 \\
 $^{234}$Th & 63.29    & 4.8     & 62.34  & 64.20  & 127 & 9 & 1   & 0.188 & 0.288 \\
 $^{226}$Ra & 187.1    & -       & 184.2  & 188.20 & 278 & 2 & 4   & 0.333 & 0.288 \\ 
$^{212}$Pb  & 238.63   & 43.3    & 236.64 & 241.97 & 306 & 1 & 5   & 0.912 & 0.218 \\
 $^{214}$Pb & 351.93   & 37.6    & 349.9  & 353.9  & 265 & 5 & 4   & 0.536 & 0.288 \\
 $^{208}$Tl & 583.2    & 84.5    & 581.20 & 585.27 & 219 & 1 & 8   & 0.801 & 0.218 \\
 $^{214}$Bi & 609.31   & -       & 607.30 & 611.35 & 225 & 2 & 0   & 0.581 & 0.288 \\
 $^{137}$Cs & 661.65   & 85.1    & 659.7  & 663.7  & 198 & 1 & 5   & 1     & 0.255 \\
 $^{228}$Ac & 911.20   & 25.8    & 909.21 & 913.26 & 182 & 2 & 9   & 0.383 & 0.181 \\
 $^{228}$Ac & 968.97   & 15.8    & 967.27 & 971.03 & 155 & 0 & 7   & 0.383 & 0.181 \\
 $^{234m}$Pa& 1001.03  & 0.84    & 999.02 &1003.08 & 188 & 0 & 166 & 0.512 & 0     \\
 $^{214}$Bi & 1120.2   & 15.1    & 1118.3 &1122.35 & 151 & 1 & 3   & 0.581 & 0.288 \\
 $^{60}$Co  & 1173.23  & 99.0    & 1171.2 &1175.2  & 176 & 0 & 0   & 1     & 0.499 \\
 $^{60}$Co  & 1332.5   & 99.0    & 1330.5 &1334.5  & 360 & 0 & 0   & 1     & 0.499 \\
 $^{40}$K   & 1460.83  & 11.0    & 1458.8 &1462.5  & 268 & 0 & 75  & 1     & 0     \\
 $^{214}$Bi & 1764.49  & 15.4    & 1762.5 &1766.5  & 85  & 0 & 1   & 0.581 & 0.288 \\
 $^{208}$Tl & 2614.5   & 99.0    & 2612.5 &2616.5  & 39  & 2 & 1   & 0.801 & 0.218 \\ \hline
\end{tabular}
\end{table*}

\subsection{Ranking without raw data}
It is often the case that the spectrum data is only 
available in image format and ranks  E, F and J 
have the advantage of performing ranking
without the raw data of the spectrum. To show this capability,
a calibrated spectrum in which isotopes were identified and marked with tags ( figure 2 of \cite{granSasso})  was analyzed using ranks F and J. 
The peak maxima and Full Width at Half Maximum reported (FWHM) were used to construct an \texttt{info} file \cite{histoGeUserManual} with 17 query ranges. The query ranges were
defined as two times the FWHM except for $^{234}$Th in which only one FWHM was used.
The positions and RV for each $\gamma$-line 
are shown in Table \ref{tab:gSasso}. For rank
F, all the $\gamma$-lines known to be present were placed within the first ten positions to explain the peak where they are found. 12 out of 17 were placed among the top three and 3 were positioned fourth. Criterion J placed the lines within the first ten positions except for $^{234m}$Pa and $^{40}$K which could not be associated with a specific decay chain in the LDB. 

\subsection{Comparison between \texttt{histoGe} and \texttt{InterSpec}\label{sec:InterSpec}}
Table \ref{tab:interSpec} shows a performance comparison between
\texttt{histoGe} and \texttt{InterSpec} from Sandia Labs \cite{interspec}, whose output is
also an ordered list.
Two spectra were analyzed: sample B of Table 
\ref{tab:waterUk} and the background spectrum of the Lumpsey
detector, at Boulby, (B-BKG) \cite{scovell2018low}. For sample B, rank H of \texttt{histoGe} outperforms
\texttt{InterSpec} while rank F is slightly better, in particular,
to identify $^{177}$Lu and the 80.185 keV photopeak of
$^{131}$I. For B-BKG, \texttt{InterSpec}
was capable of ranking correctly more $\gamma$-lines at first position in
comparison to the best result of \texttt{histoGe}, however, there
is one peak in which it fails completely ($^{208}$Tl
at 510.77 eV) because it is confused with $e-e+$ annihilation line. So,
if $\Delta$ is defined as the position difference
between \texttt{InterSpect} and the \texttt{histoGe}'s  best result, it can be seen that $\sum \Delta > 0$, which means that
\texttt{histoGe} had a better overall performance when both, sample B and B-BKGD are considered.

\begin{table*}[!ht]
\centering

\caption{\centering \textsc{\label{tab:interSpec} Comparison of the results obtained between \texttt{histoGe} vs. \texttt{InterSpec} (IS). $\Delta$ means the difference of position of \texttt{InterSpec} minus position of \texttt{histoGe}. A positive $\Delta$ favors \texttt{histoGe}. Due to $^{214}$}Bi \textsc{has more than 2 hundreds of peaks the information to identify it clearly is incomplete making harder the identification for \texttt{histoGe}. $^\dagger$Range had to be made wider so IS could make identification. $^{\dagger\dagger222}$}Rn, \textsc{$e-e+$ and $^{214}$}Pb \textsc{were identified at positions 0, 1 and 2, respectively and both isotopes were found in the background.}} 
\resizebox{!}{0.5\columnwidth}{
\begin{tabular}{|c|c|c|c|c|c|c|c|c|c|}\hline
\multicolumn{1}{|c|}{}&\multicolumn{2}{|c|}{Known isotope} & \multicolumn{2}{|c|}{Range}& \multicolumn{4}{|c|}{Position}  & \multicolumn{1}{|c|}{} \\
\hline
 spectrum & isotope & Eg [keV] & Emin & Emax   & F & H & H+ & IS &$\Delta$  \\ \hline \hline

\multirow{6}{*}{Sample B} & $^{177}$Lu & 112.94& 111.1131 & 115.1425  & 1 & 0& 0& 1& 1\\
                   & $^{177}$Lu & 208.36 & 206.8 & 209.8  & 3 & 0& 0 &   2 & 2\\
                   & $^{131}$I  & 80.18 & 79.3  & 82.0   & 4 &  0 & 0 & 6 & 6\\
                   & $^{131}$I  & 284.30 & 282.5  & 285.5   & 0 & 0 & 0 &  0 & 0\\
                   & $^{131}$I  & 364.48 & 362.51 & 366.1  & 0 & 0& 0 &    0 & 0 \\
                   &$^{131}$I   & 636.98 & 634.64 & 639.0  & 0 & 0& 0 &    0 & 0\\
                   &$^{40}$K   & 1460.83 & 1457.0 & 1463.53  & 0 & 0& 0  &  0 &0\\ \hline
\multirow{6}{*}{B-BKG} & $^{214}$Bi & 609.31 & 607.7 & 611.3 & 2 & 1 & 2 & 0 & -1\\
                       & $^{214}$Bi & 768.35 & 767.80 & 769.82 & 1 & 0  & 0 & 0 & 0\\
                       & $^{214}$Bi & 1120.28 & 1119.10 & 1122.55 & 1 & 1  & 1 & 0 & -1\\
                       & $^{214}$Bi & 1764.49 & 1763.35 & 1767.20 & 1 & 1 & 1& 0 & -1\\
                       & $^{214}$Bi & 2204.21 & 2203.50 & 2207.16 & 0 & 0 & 0 & 0 & 0 \\
                       & $^{40}$K & 1460.83 & 1459.5 & 1462.8 & 0 & 0 & 0 & 0 & 0 \\
                        & $^{228}$Ac & 911.20 & 909.7 & 913.16 & 1 & 1 & 2 & 0 & -1  \\
                       & $^{228}$Ac & 968.97 & 967.57 & 971.23 & 0 & 0 & 0 & 0 & 0  \\
                       & $^{228}$Ac & 1588.19 & 1587.5 & 1589.77 & 0 & 0 & 0 & 0 & 0  \\
                        & $^{212}$Pb & 47.91 & 45.1 & 47.92 & 3 & 0 & 0 & 0$^\dagger$ & 0  \\
                        & $^{212}$Pb & 238.63 & 237.00 & 239.95  & 0 & 0 & 0& 0 & 0  \\
                       & $^{212}$Pb & 300.087 & 298.9 & 301.3 & 0 & 0 & 0 & 0 & 0  \\
                        & $^{208}$Tl & 510.77 & 509.82 & 512.16  & 2 & 3$^{\dagger\dagger}$ & 0 & 26 & 26  \\
                        & $^{208}$Tl & 583.19 & 581.8 & 584.89  & 1 & 2 & 1 & 0 & -1  \\
                        & $^{208}$Tl & 860.56 & 859.2 & 862.5   & 0 & 0 & 0 & 0 & 0  \\
                        & $^{208}$Tl & 2614.53 & 2613.65 & 2617.31 & 1 & 2 & 0 & 0 & 0  \\
                        & $^{214}$Pb & 241.99 & 240.3 & 243.2  & 1 & 0 & 0 & 4 & 4  \\
                        & $^{214}$Pb & 295.22 & 293.76 & 296.3   & 0 & 1 & 0 & 0 & 0  \\
                        & $^{214}$Pb & 351.93 & 350.52 & 353.3 & 2 & 3 & 0 & 0 & 0  \\ \hline

\end{tabular}
}
\end{table*}

The paradigm by which \texttt{histoGe} identifies
the isotopes is completely different in comparison
to \texttt{InterSpec}'s. \texttt{histoGe} performs better when 
more photopeaks of an isotope are
observed in the spectrum, however, sometimes \texttt{InterSpec}
works better when individual peaks are considered in
such a way that even identification of 2 unrelated
peaks could make \texttt{InterSpec} fail the isotopes'
identification unlike \texttt{histoGe} which is able to
manage contextually many isotopes per run.

\subsection{Computational cost}

The computational cost of criteria F and H is due to querying the LDB, calculating the RV, and sorting all the
$\gamma$-lines. A Monte Carlo analysis was performed to estimate the 
time used for these process.
The procedure followed to implement 
this test is described next: \texttt{.info} files were randomly 
constructed with a variable number of $\gamma$-lines per file, and then, 
ranked.  The center of the energy range was chosen randomly between $200$ 
keV and $2500$ keV per each $\gamma$-line and its width was calculated 
using a distributed normally random numbers  with $\mu=0$ and $\sigma=5$. 
The minimum energy range is $1$ keV to avoid small intervals that could
contain few $\gamma$-lines. This procedure was repeated 100 times. The 
computer used to execute this analysis has a AMD Ryzen 7 processor 4800H, 16 GB of RAM and a SSD. A program was made to measure the execution time of
each call to \texttt{histoGe} using those \texttt{.info} files generated randomly. For H criterion, an arbitrary spectrum was chosen considering that its energy range is larger than the query ranges. Figure 
\ref{fig:MonteCarlo} shows that the execution time follows a non-linear 
relationship in which increasing the number of peaks by 10 does not even 
get the execution time doubled. These results show that \texttt{histoGe} 
has reasonable ($\geq 20s$) execution times for real spectra in which the number of 
$\gamma$-lines do not exceed a hundred peaks. As expected, H is slower than F, but for a low number of peaks, there is not a considerable difference between them.

On the other hand, note that processing a spectrum form  peak identification  using the PDA through the
\texttt{histoGe}'s peak finder tool to $\gamma$-line identification using some criteria could take a few minutes,
however, due to the peak finding algorithms are not fully accurate more time could be required
to adjust the query ranges in the \texttt{info} file by a trained spectroscopist.  

\pgfplotsset{compat=1.17,every axis/.append style={font=\small,tick style={thin}}}
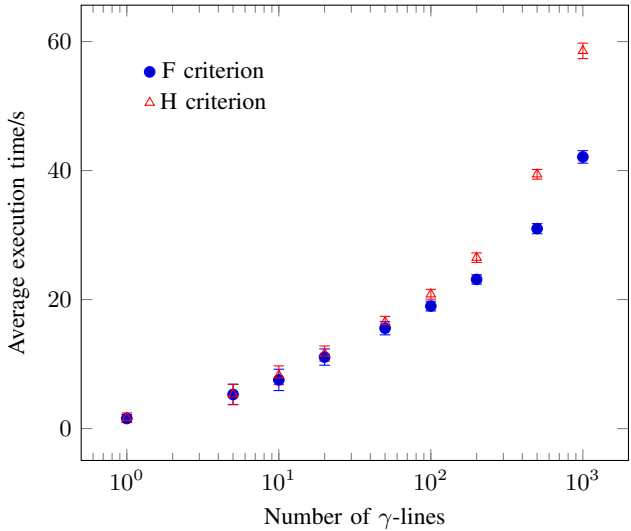
\begin{figure}[htb]
    \centering
    \begin{tikzpicture}
\begin{semilogxaxis}[
xlabel={Number of $\gamma$-lines},
ylabel={Average execution time/s},
legend style={
draw=none,
at={(0.1,0.9)},
anchor=north west},
]
\addplot+[only marks, mark=*,color=blue] 
  plot[error bars/.cd, y dir=both, y explicit]
table[x={x},y={y},y error plus expr=\thisrow{ydev},y error minus expr=\thisrow{ydev}]{\mytable};

\addplot+[only marks, mark=triangle,color=red] 
  plot[error bars/.cd, y dir=both, y explicit]
  table[x={x},y={z},y error plus expr=\thisrow{zdev},y error minus expr=\thisrow{zdev}]{\mytable}; 
\legend{F criterion,H criterion}

\end{semilogxaxis}
\end{tikzpicture}
\caption{Average and error bars (2$\sigma$) of the execution time of criteria F (blue) and H (red) vs. the number $\gamma$-lines per info file.}
\label{fig:MonteCarlo}
\end{figure}

\section{\label{sec:conclusions}Conclusions}

\texttt{histoGe} is a tool
for the identification of 
peaks in a $\gamma$ spectrum using the information in a large 
database containing 92,453 gamma lines and 2,200 radioactive 
nuclides. The program implements different criteria to rank 
and sort candidate isotopes to explain the presence of peaks 
in a spectrum with a philosophy inspired by that of a search 
engine.

Five different criteria (\ref{sec:crit}) were presented and
their performances were compared according to their ability
to identify the $\gamma$-lines from a suite of sealed
calibrated radioactive gamma sources via an \textit{ad hoc}
defined metric. Two of the methods stood out in performance
under this test: one based on the use of Relative Emission
Probabilities (F), and one using fuzzy logic (H) which
combine information used in other criteria. An enhanced
version of the latter (H+) where $\gamma$-lines with
relatively small REPs are discarded achieved
$\sim 95\%$ efficiency to identify the isotopes in a set of 
sealed radioactive sources. Under the same assumptions,
\texttt{InterSpec} was also tested giving an efficiency
of $70\%$, demonstrating that under this conditions
\texttt{histoGe} performed better.
$^{241}$Am was a hard isotope to identify, as only a small 
fraction of its $\gamma$-lines are typically visible in spectra.
The score increases quite noticeably
when $^{241}$Am is not considered. 

\texttt{histoGe} was used to identify the $\gamma$-lines in
three arbitrary samples: water with lead (sample A), London tap
water (sample B) and the rock of LabChico's site (sample C).
For sample A, a $\gamma$-line of $^{210}$Pb was identified with
accuracy but X-rays were not because the database has not
complete information about X-rays. For sample B, criterion H
got the best score when $^{177}$Lu, $^{131}$I and
$^{40}$K were identified. For sample C, $^{214}$Bi was hardly
identified by rank H due to it having 290 $\gamma$-lines but rank
F performed better and H+ identified it with accuracy.

An advantage of \texttt{histoGe} is that it is capable
of making isotope identification using only the ranges
of the photopeaks without the raw data of the spectrum.
Criteria E, F, and J are capable of doing this. Then,
a published spectrum was ranked with criterion F, it was
capable to identify all of them within the 10 first positions,
in fact most of them were among the first three places.
Chain rank (\ref{sec:CE2}) identified all $\gamma$-lines
in the top ten places except  $^{40}$K and $^{234m}$Pa
because there is not a decay chain associated with them.
This result motivates further research about how to
improve this rank. 

\texttt{histoGe}'s results of identification of
two samples were compared to those obtained with
\texttt{InterSpec} of SandiaLabs. In general, the best
result of \texttt{histoGe} outperformed \texttt{InterSpec}.
Besides, from the results presented in section \ref{sec:InterSpec} and figure \ref{fig:scores}, it can be seen that \texttt{histoGe} is capable to make identification when multiple related or unrelated $\gamma$-lines are given at the same time, on the contrary, \texttt{InterSpec} cannot identify accurately unrelated
peaks but is highly accurate in individual
identification. These findings may change if
both are tested under different conditions.
Using the ideas presented in this work,
a \texttt{histoGe} and \texttt{InterSpec} could be combined
to get an even more accurate one.

From the Monte Carlo study, it was found that the computational cost for
real spectra is affordable. The overall results have shown that
\texttt{histoGe} is a tool capable to make reliable identification
of isotopes through $\gamma$-spectroscopy with results comparable
or even better than other similar software, which makes it able
to be applied in teaching research and industrial applications. However, in its current state, we recommend that a trained spectroscopist analyze the results obtained with \texttt{histoGe} or \texttt{InterSpec} to get a better interpretation.

\section*{Acknowledgment}

The authors would like to thank Prof. Peter Ekström (Lund University) and Prof. Dirk Rudolph (Lund University) for giving permission to use their
database. We acknowledge the help of the Radiological Safety Unit of ICN-UNAM.

\bibliographystyle{IEEEtran}
\bibliography{main}

\end{document}